\documentclass[showpacs,amsmath,amssymb,12pt]{revtex4}
\usepackage{graphicx}
\newcommand{\be}{\begin{equation}}
\newcommand{\ee}{\end{equation}}
\newcommand{\bea}{\begin{eqnarray}}
\newcommand{\eea}{\end{eqnarray}}
\newcommand{\bean}{\begin{eqnarray*}}
\newcommand{\eean}{\end{eqnarray*}}

\begin{document}
\title{Role of t-channel meson exchange in S-wave $\pi N$ and $K N$ scattering}
\author{Feng-Quan Wu $^{1,2}$}
\email{wufq@mail.ihep.ac.cn}
\author{Bing-Song Zou$^{1,3}$}
\email{zoubs@mail.ihep.ac.cn} \affiliation{$^{1}$ Institute of High
Energy Physics, Chinese Academy of Sciences,P.O.Box 918(4), Beijing
100049, China\\$^{2}$ National Astronomical Observatories, Chinese
Academy of Sciences, Beijing 100012, China \\$^3$ Theoretical
Physics Center for Science Facilities, CAS, Beijing 100049, China }
\date{\today}
\begin{abstract}
The low-energy S-wave $\pi N$ and $K N$ scatterings are studied by
the K-matrix approach within the meson exchange framework. The
t-channel meson exchanges, especially $\rho$ and $\sigma$ exchanges,
are found to play a very important role in these two processes. The
t-channel $\rho$ exchange determines the isospin structure of the
scattering amplitudes,  it gives attractive force in the low isospin
state but repulsive force in the high isospin state.
 The t-channel $\sigma$ exchange gives a very large
contribution in these two processes,  while it is negligible in
 meson-meson S-wave scatterings.

\end{abstract}
 \maketitle

The $\pi N$ and $K N$ scatterings are crucial sources of information
about strong interaction.  The wealth of accurate data and the
richness of structures shown by them provide an excellent but also
challenging testing ground for many models.

The $\pi N$ interaction is one of most fascinating hadron-hadron
interactions for several reasons. Firstly, it is one of main sources
of information about the baryon spectrum. We know that most of
experimental information about the mass, width, and decay of baryon
resonances is extracted from partial wave analysis of $\pi N$
scattering data.  Secondly, the $\pi N$ scattering have accumulated
a large amount of rich and accurate data.  It provides a unique
place for testing various theoretical approaches, such as
meson-exchange model and chiral perturbation theory.  Finally, the
$\pi N$ interaction is an important ingredient in many other
hadronic reactions, such as meson production in nucleon-nucleon
collision.

For the $K N$ scattering,  it is worthy to mention that kaons have
two properties which make them unique projectile for investigating
nuclear structure. Firstly, they can transfer a new degree of
freedom to nucleus, and secondly, in contrast to pions they come in
two forms, kaons ($K$) and antikaons ($\bar{K}$) which differ
substantially in their interaction with nucleus. Because of a
strangeness quantum number conserved in strong interactions, the
interaction between $K$ and nuclei is rather weak. Consequently,
the $K$ meson is a suitable probe for investigating the interior
region of nuclei.

Over the past ten years, in a series of papers, the J\"{u}lich
group has investigated the $\pi N$ \cite{julich pi N} and $K N$
\cite{julich K N} scattering in the meson exchange framework using
time-ordered perturbation theory.  At the same time, much  other
efforts have also been devoted to the study of the $\pi N$ \cite{pi
N} and $K N$ \cite{K N} scattering.

As well known,  the conventional t-channel meson exchange final
state interaction mechanism \cite{rho-exchange,Wu:2003wf} can give
consistent explanation for the broad $\sigma$ near $\pi\pi$
threshold,  the broad $\kappa$ near $K\pi$ threshold,  the narrow
$f_0(980)$ peak near $\bar KK$ threshold, the narrow structure near
$\bar pp$ threshold \cite{FSI},  and ``b1 puzzle" in the $J/\Psi \to
\omega \pi \pi$ decay\cite{Wu:2006tb}.   Since t-channel meson
exchange plays such an important role in low-energy hadron physics,
it is necessary to study the role of  t-channel meson-exchange
contribution in the S-wave $\pi N$ and $K N$ interaction.

The purpose of the present paper is to reveal the t-channel meson
contribution in S-wave  $\pi N$ and $K N$ scatterings,
consequently, to give a consistent qualitative explanation for some
familiar S-wave low-energy meson-meson and meson-baryon scatterings
in the meson exchange framework.
 The role of
t-channel $\rho$ in S-wave meson-meson scattering, such as $\pi \pi$
and $\pi  K$ scattering, has been studied in Ref.\cite{Li}. The
S-wave phase shift analysis shows t-channel $\rho$ gives attractive
force in isospin I=0 channel for  $\pi \pi$ scattering (or in
I=$1\over 2$ channel for  $\pi K$ scattering), but repulsive forces
in isospin I=2 channel for $\pi \pi$ scattering (or in I=$3\over 2$
channel for  $\pi K$ scattering).  The t-channel $\sigma$ exchange
gives a very small contribution in $\pi \pi$ and $\pi  K$ S-wave
scattering, and can be neglected in these processes.  In this paper,
we will continue to discuss the role of t-channel meson, such as
$\rho$, $\phi$ and $\sigma$ exchanges,  in the $\pi N$ and $K N$
scattering.

The paper is organized as follows. In the next section,   we
present the method and formulation for calculation.  The numerical
results and discussions are given in section II.



\section{Method and formulation}
\begin{figure}[htbp]
\begin{minipage}[t]{0.45\textwidth}
\includegraphics[scale=0.7]{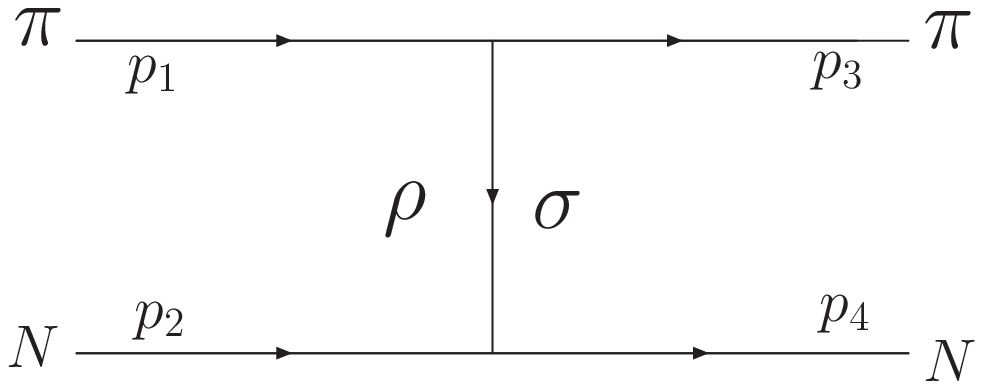}
\end{minipage}
\hfill
\begin{minipage}[t]{0.45\textwidth}
\includegraphics[scale=0.7]{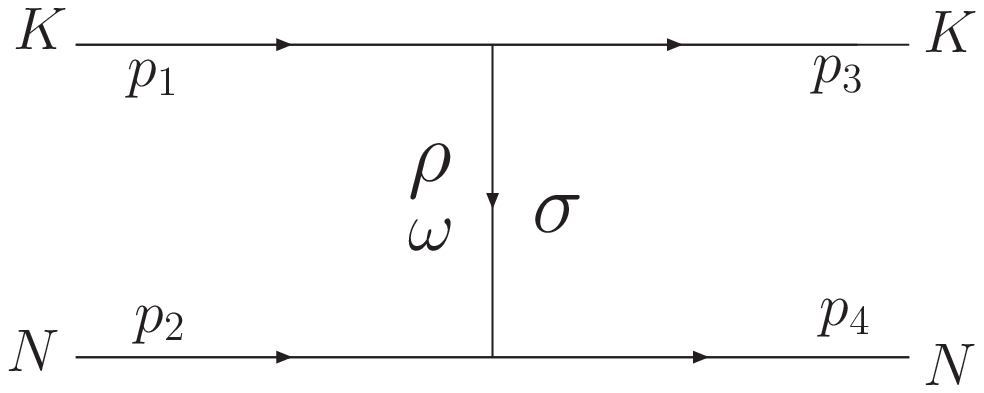}
\end{minipage}
\caption{Tree diagrams for $\pi N$ and $K N$ scattering with single
meson exchange. }
 \label{fig:piNKN}
\end{figure}

The Feynman diagrams for relevant processes are depicted in
Fig.\ref{fig:piNKN}. As the first step the processes arising from
the s-channel resonances ($N^*$, $\Delta^*$), u-channel baryon ($ N,
N^*, \Delta, \Lambda, \Sigma$) exchanges, and higher-order diagrams
are not considered. The main purpose of this work is examining the
t-channel exchange contributions.


We start with the baryon-baryon-meson couplings. The interaction
lagrangians are listed as follows:
  \bea
{\cal L}_{NN\rho} &=&-g_{NN\rho}
\bar{\Psi}[\gamma^{\mu}-\frac{\kappa_{\rho}}{2m_N}\sigma^{\mu\nu}\partial_{\nu}]\vec{\tau}\vec{\rho}_{\mu}
\Psi,   \\
  {\cal L}_{NN\omega} &=& -g_{NN\omega} \bar{\Psi}\gamma^{\mu}\omega_{\mu}
  \Psi,  \\
  {\cal L}_{NN\sigma} &=& -g_{NN\sigma} \bar{\Psi} \Psi \sigma,
 \eea

For the pseudoscalar-pseudoscalar-vector coupling, we use the
SU(3)-symmetric Lagrangian \cite{lagrangian,Li}. \be
\mathcal{L}_{PPV}=-\frac{1}{2}iG_V\text{Tr}([P,\partial_\mu
P]V^\mu), \ee where $G_V$ is the coupling constant, $P$ is the
$3\times3$ matrix representation of the pseudoscalar meson octet,
$P=\lambda^aP^a, a=1,\ldots,8$ and $\lambda^a$ are the $3\times3$
generators of SU(3),
\be%
\label{eq:p matrix} P = \sqrt{2} \left(
\begin{array}{ccc}
\frac{1}{\sqrt{2}}\pi^0 + \frac{1}{\sqrt{6}}\eta_8 & \pi^+ & K^+\\
\pi^- & -\frac{1}{\sqrt{2}}\pi^0 + \frac{1}{\sqrt{6}}\eta_8 & K^0\\
K^- & \bar{K}^0 & - \frac{2}{\sqrt{6}}\eta_8
\end{array}
\right).
\ee%

A similar definition of $V_{octet}$ is used for the vector meson
octet. In the large $N_c$ limit, the octet and singlet vector mesons
can be combined into a single ``nonet" matrix $V$
\cite{Jenkins:1995vb},
\be%
\label{eq:V matrix} V = V_{octet}+
\sqrt{\frac{2}{3}}\omega_0=\sqrt{2} \left(
\begin{array}{ccc}
\frac{1}{\sqrt{2}}\rho^0 + \frac{1}{\sqrt{2}}\omega & \rho^+ & K^{*+}\\
\rho^- & -\frac{1}{\sqrt{2}}\rho^0 + \frac{1}{\sqrt{2}}\omega & K^{*0}\\
K^{*-} & \bar{K}^{*0} & \phi
\end{array}
\right),
\ee%
where the standard $\omega-\phi$ mixing is assumed.

In the Gell-Mann representation, the Lagrangian can be expressed as
\be \mathcal{L}_{PPV}=2G_Vf_{abc}P^a\partial_\mu P^bV^{c\mu}, \ee
where $f_{abc}$ are the antisymmetric structure constants of SU(3).
For example, \bea \mathcal{L}_{\rho\pi\pi} &=& 2 G_V [(p_{\pi^+}^\mu
-p_{\pi^-}^\mu )\rho^0_{\mu} + (p_{\pi^-}^\mu -p_{\pi^0}^\mu
)\rho^+_{\mu} +
(p_{\pi^0}^\mu -p_{\pi^+}^\mu )\rho^-_{\mu}] \\
  \mathcal{L}_{\rho K \bar{K}} &=& G_V [(p_{K^+}^\mu - p_{K^-}^\mu)
  \rho^0_\mu + (p_{\bar{K}^0}^\mu - p_{{K}^0}^\mu) \rho^0_\mu
] + \nonumber \\ && \sqrt{2}~ G_V (p_{K^0}^\mu - p_{K^-}^\mu)
\rho^+_\mu + \sqrt{2}~ G_V (p_{K^+}^\mu - p_{\bar{K}^0}^\mu)
\rho^-_\mu  \label{psi rho k
k}     \\
\mathcal{L}_{\omega K \bar{K}} &=&  G_V [(p_{K^+}^\mu -
p_{K^-}^\mu)
  \omega_\mu + (p_{{K}^0}^\mu - p_{\bar{K}^0}^\mu) \omega_\mu
] \eea

The employed pseudoscalar-pseudoscalar-scalar coupling are
  \bea
{\cal L}_{\pi\pi\sigma} &=&  \frac{g_{\pi \pi\sigma }}{2 m_{\pi}}
\partial_{\mu} \vec{\pi} \partial^{\mu} \vec{\pi} \sigma,\\
{\cal L}_{K K\sigma} &=&  \frac{g_{ K K \sigma}}{2 m_{\pi}}
\partial_{\mu} \bar{K} \partial^{\mu} {K} \sigma,
 \eea
where
 $K\equiv\left(\begin{array}{c}K^{+}\\K^{0}\end{array}\right)$,
 and $\bar{K}\equiv\left(\begin{array}{cc}K^{-} &
 \bar{K^{0}}\end{array}\right)$ , and  $g_{\pi \pi\sigma }= g_{ K K
 \sigma}$
 due to flavor SU(3) symmetry.

Using these Lagrangians,  we are able to construct following
amplitudes corresponding to the diagrams in Fig.\ref{fig:piNKN},
respectively:
 \bea
  T_{\pi N}^{\rho} &=&
   - \frac{G_V g_{NN\rho}}{8 \pi}
      \frac{\mbox{IF}}{(p_1-p_3)^2 - m_\rho^2} \bar{u}(p_4)[(\not\! p_1 + \not\! p_3) -
      \frac{\kappa_{\rho}}{2m_N}(-\not\! p_1 \not\! p_3 + \not\! p_3 \not\!
      p_1)]u(p_2) , \\
 T_{\pi N}^{\sigma} &=& - \frac{g_{NN\sigma}g_{\pi\pi\sigma}}{16 \pi m_\pi }
       \frac{\mbox{IF}}{(p_1-p_3)^2-m_\sigma^2}\bar{u}(p_4)u(p_2) p_1 \cdot p_3,\\
  T_{K N}^{\rho}
     &=&    - \frac{G_V g_{NN\rho}}{16 \pi}
      \frac{\mbox{IF}}{(p_1-p_3)^2 - m_\rho^2} \bar{u}(p_4)[(\not\! p_1 + \not\! p_3) -
      \frac{\kappa_{\rho}}{2m_N}(-\not\! p_1 \not\! p_3 + \not\! p_3 \not\!
      p_1)]u(p_2), \\
 T_{K N}^{\omega} &=&  - \frac{G_V g_{NN\omega}}{16 \pi}
      \frac{\mbox{IF}}{(p_1-p_3)^2 - m_\omega^2} \bar{u}(p_4)(\not\! p_1 + \not\!
      p_3)u(p_2),\\
 T_{K N}^{\sigma} &=&   - \frac{g_{NN\sigma}g_{K K\sigma}}{16 \pi m_\pi }
     \frac{\mbox{IF}}{(p_1-p_3)^2-m_\sigma^2}\bar{u}(p_4)u(p_2) p_1 \cdot p_3,
 \eea
where  $T_{\pi N}^{\rho}$ denotes t-channel $\rho$ meson exchange
amplitude for $\pi N$ scattering,  and $\mbox{IF}$ is isospin factor
listed in Table \ref{table:IF} for various processes.
\begin{table}[htbp]
\begin{tabular}{|c|c|c|c|}
 \hline\hline
  Process & Exch. part. & ~~~I~~~  & ~~~IF~~~ \\
  \hline
  $\pi N \to \pi N$ & $\rho$ & $\frac{1}{2}$ & -2 \\
                     &       & $\frac{3}{2}$ & 1  \\
                    & $\sigma$ & $\frac{1}{2}$ & 1  \\
                      &       & $\frac{3}{2}$ & 1  \\
   $K N \to K N$ & $\rho$ & 0  & -3 \\
                     &    & 1   & 1  \\
                      & $\omega$, $\sigma$& 0  & 1 \\
                     &    & 1   & 1  \\
 \hline\hline
\end{tabular}
\label{table:IF}
 \caption{Isospin factors for the two processes.}
\end{table}

Usually, hadronic form factors should be applied to the
baryon-baryon-meson vertices because of the inner quark-gluon
structure of hadrons.  Due to the difficulties in dealing with
nonperturbative QCD hadron structure, the form factors are commonly
adopted phenomenologically.  The most commonly used form factor for
baryon-baryon-meson vertices in t-channel is
\begin{equation}
F(\Lambda, q)=\frac{\Lambda^2-m^2}{\Lambda^2-q^2},
\end{equation}
where $m$ and $q$ are the mass and  four-momentum of  intermediate
particle, respectively, and $\Lambda$ is the so-called cut-off
momentum that can be determined by fitting the experimental data.
The parameters of model are listed in Table \ref{table:param}.

\begin{table}[htbp]
\begin{center}
\begin{tabular}{c c c c c}
\hline\hline
Vertex & Exchange particle & Coupling constant & Ref. & Cutoff $\Lambda$ [MeV] \\
\hline
$NN\rho$ & $\rho$  & $\frac{g^2_{NN\rho}}{ 4\pi} =0.84$ & \cite{Janssen:1996kx}& {\bf 1500} \\
& & $\kappa=6.1$ & \cite{Janssen:1996kx}&  \\
$NN\omega$ & $\omega$ & $\frac{g^2_{NN\omega}}{ 4\pi} =7.563$& \cite{julich K N}& {\bf 1500} \\
$NN\sigma$ & $\sigma$ & $\frac{g^2_{NN\sigma}}{ 4\pi} =13$&\cite{Durso:1980vn} & {\bf 2000} \\
& & $( m_\sigma$=0.65 GeV ) &  &  \\
$\pi\pi\sigma$ & $\sigma$ & $\frac{g^2_{\pi\pi\sigma}}{4\pi}
=0.25$&\cite{Krehl:1997kg} & {\bf 2000} \\
$\pi \pi \rho$ & $\rho$ & $\frac{G_V^2}{8\pi}
=0.364$&\cite{Wu:2003wf} & {\bf 1500} \\
 $K \bar{K}
\rho$ & $\rho$ & $\frac{G_V^2}{8\pi} =0.364$&\cite{Wu:2003wf} &
{\bf 2000} \\
 $K \bar{K}
\omega$ & $\omega$ & $\frac{G_V^2}{8\pi} =0.364$&\cite{Wu:2003wf} &
{\bf 1500} \\
 $K \bar{K}
\sigma$ & $\sigma$ & $\frac{g^2_{K K \sigma}}{4\pi} =0.25$
&\cite{Krehl:1997kg} &
{\bf 2000} \\
\hline \hline
\end{tabular}
\caption{Parameters of the model. Free parameters are given in
boldface.}
\label{table:param}
\end{center}
\end{table}

In order to isolate the s-wave contribution,  we have to perform a
partial wave decomposition of the amplitudes. Our normalization is
described as follows: the differential cross-section in the
centre-of-mass system is given by the Lorentz-invariant matrix
element $\mathcal{M}$ as
 \bea \frac{d \sigma}{d \Omega}= \frac{1}{64 \pi^2 s}
 |\mathcal{M}|^2,
 \eea
where s is the invariant mass squared.  The relation between the $T$
matrix and the $\mathcal{M}$ matrix is
 \bea T=\frac{1}{16\pi} \mathcal{M}= g(s,\theta) + i h(s,\theta)
 {\mbox{\boldmath{$\sigma$}} \cdot \bf \hat{n}},
 \eea
here  ${\bf  \hat{n}}$ is the  unit vector normal to the scattering
plane.  The amplitude $T$ can be expanded in
orbital angular momentum $l$ and total angular  momentum $J$,
 \bea T= \sum_l (2l +1)[f_{l+} \hat{Q}_{l+} + f_{l-}  \hat{Q}_{l-}]
 P_l(\cos \theta),
 \eea
where $P_l(x)$ are Legendre polynomials and $\hat{Q}_{l \pm}$ are
the corresponding projection operators for $J=l\pm \frac{1}{2}$,
 \bea \hat{Q}_{l+}=\frac{l+1+ {\mbox{\boldmath{$l$}} \cdot \mbox{\boldmath{$\sigma$}}}}{2l+1},
 ~~~~~\hat{Q}_{l-}=\frac{l- {\mbox{\boldmath{$l$}} \cdot \mbox{\boldmath{$\sigma$}}}}{2l+1}.
 \eea

Then one obtains
 \bea g(s,\theta) &=& \sum_{l}[(l+1)f_{l+}+lf_{l-}]P_l(\cos
 \theta),  \\
  h(s,\theta) &=& \sin \theta \sum_{l}[f_{l+} - f_{l-}]P^{'}_l(\cos
  \theta),
 \eea
  where $P^{'}_l(x)=(d/dx)P_l(x)$. For each partial wave $a=l\pm$,
  the phase shift is related to $f_a$ by
  \bea f_a= \frac{\eta_a e^{2i\delta_a}-1}{2 i \rho}.
  \eea

\section{Numerical results and discussions}

Having described our model, we turn now to compare its results to
the experimental data. First of all, we discuss the parameters that
enter into our model calculation. All of the coupling constants have
been taken from other sources as listed in Table \ref{table:param}.
We have varied only boldface printed values in Table
\ref{table:param}.   It is worth to mention that we take the
coupling constant of vertex $N N \omega$, $g_{N N \omega}$, as its
normal value obtained by assumption of SU(3) symmetry related to the
empirical $N N \pi$ coupling.  We have not increased $g_{N N
\omega}$ and $g_{K K \omega}$  as done in Ref. \cite{julich K N}. As
discussed in Ref.\cite{julich K N},  this increased
$\omega$-exchange leads to additional repulsion in P- and higher
waves too, which, as a whole, seems not be favoured by the empirical
data, especially in the $P_{03}$ and $P_{11}$ channels.

\begin{figure}[htbp]
\includegraphics[scale=1.5]{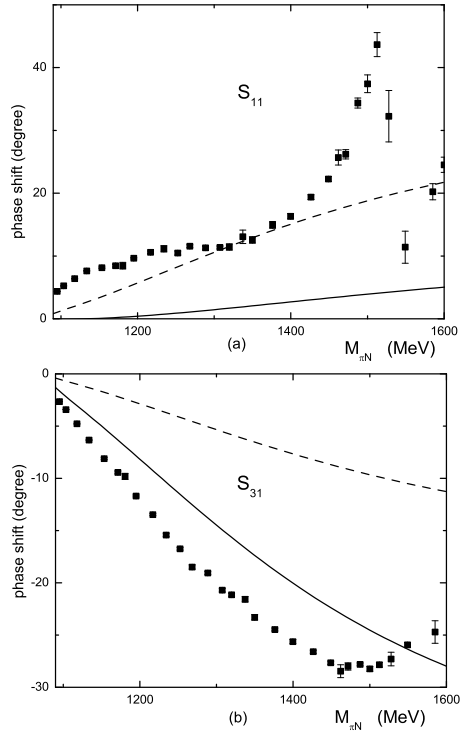}
\caption{The S-wave $\pi N$ phase shifts. The dashed curves show
the results from only $\rho$ exchange.  The solid curves represent
the contribution of t-channel $\rho$ and $\sigma$ exchanges
together. The data are from the phase-shift analysis SM95
\cite{sm95}.}
 \label{fig:piNs}
\end{figure}

From the formalism given above and using the parameters listed in
Table \ref{table:param},  we obtain S-wave $\pi N$ phase shifts as
shown in Fig. \ref{fig:piNs}.  The phase shifts given by t-channel
$\rho$ exchange (dashed curves) agree with the trend of
$I=\frac{1}{2}$ and $I=\frac{3}{2}$ experimental data. It is quite
clear that t-channel $\rho$ gives attractive force in $I=\frac{1}{
2}$ channel, but repulsive force in $I=\frac{3}{2}$ channel, which
are consistent with the results  we obtain in $\pi \pi$ and $\pi K$
scattering as mentioned above, namely it gives attractive force in
the low isospin state but repulsive force in the high isospin state.
The t-channel $\sigma$ exchange provides a very large contribution
in contrast to its negligible effect in meson-meson scatterings, it
gives repulsive force both in $I=\frac{1}{2}$ and $I=\frac{3}{2}$
channel. Using the Dalitz-Tuan method \cite{Li,Wu:2003wf}, we
combine the contribution of $\rho$ and $\sigma$ exchanges together.
The results are shown in solid curves. It is found that the $\pi N$
scattering phase shifts are only in qualitative agreement with the
experimental data. Evidently, the discrepancies are primarily due to
the presence of resonances in these partial waves,  which are not
yet included in our calculation.  As discussed in Ref.\cite{julich
pi N}, the low energy $S_{11}$ phase shift they obtain in their
model even changes its sign when all couplings to the $N^*(1650)$
resonance are switched off.  Therefore, including the resonances
into calculation seems to be very important to obtain quantitative
agreement with the S-wave $\pi N$ scattering experimental data.

\begin{figure}[htbp]
\includegraphics[scale=1.5]{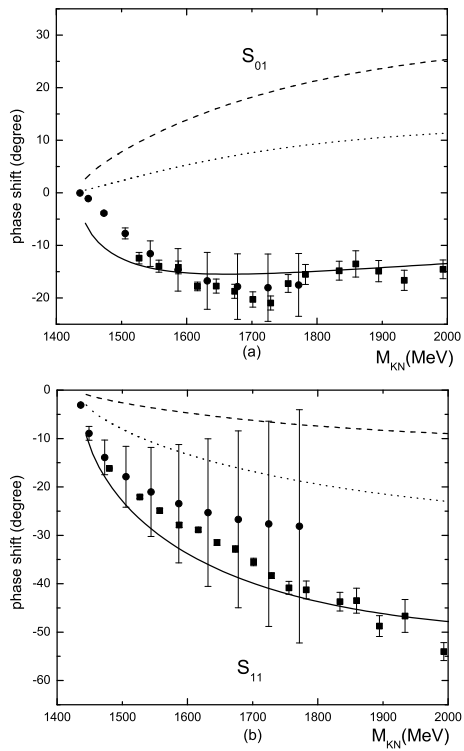}
\caption{The S-wave $K N$ phase shifts. The dashed curves show the
results from only $\rho$ exchange.  The dotted curves represent the
contribution of t-channel $\rho$ and $\omega$ exchanges together.
The solid curves involve the t-channel $\rho$, $\omega$ and
$\sigma$ exchanges totally. Experimental phase shifts are taken from
Ref.\cite{sp} (circles) and \cite{fa84} (squares).}
 \label{fig:KNs}
\end{figure}

Now, let us turn our attention to $K N$ scattering phase shifts in
Fig. \ref{fig:KNs}.  The dashed curves show the results from only
$\rho$ exchange.  Similar to $\pi \pi$, $\pi K$ and $\pi N$
scatterings, the t-channel $\rho$ gives attractive force in $I=0$
channel, but repulsive force in $I=1$ channel.  The t-channel
$\omega$ and $\sigma$ exchanges give repulsive forces both in $I=0$
and $I=1$ channels. The dotted curves represent the contribution of
t-channel $\rho$ and $\omega$ exchanges together. The solid curves
involve the t-channel $\rho$, $\omega$ and $\sigma$ exchanges
totally.  The result agrees with the experimental data quite well up
to 2.0 GeV without need to increase the value of $g_{N N \omega}$
and $g_{K K \omega}$.  From Fig. \ref{fig:KNs},  it is obvious that
$\sigma$ gives the largest contribution among these three mesons.

In summary, we study the low-energy S-wave $\pi N$ and $K N$
scatterings in the meson exchange model framework and using the
K-matrix approach. In view of the result of this paper, plus
 our previous papers in  $\pi \pi$ and $\pi K$ scatterings, we draw
the conclusion that the t-channel $\rho$ exchange determines the
isospin structures of low-energy S-wave phase shifts of these
scatterings because of its absolutely different isospin factors in
the different isospin states.  The t-channel $\rho$ exchange gives
attractive force in the low isospin state but repulsive force in
the high isospin state. In the $\pi \pi$ and $\pi K$ scatterings,
the contribution of  t-channel $\sigma$ exchange is very small,
even can be neglected.   But in the $\pi N$ and $K N$ scattering,
it  gives a very large contribution.

\section*{ACKNOWLEDGEMENTS}
One of us (F.Q.W.) would like to thank Bo-Chao Liu, Ju-Jun Xie for
valuable comments and discussions during the preparation of this
paper. This work is partly supported by the National Nature Science
Foundation of China under grants Nos. 10435080, 10521003 and by the
Chinese Academy of Sciences under project No. KJCX3-SYW-N2.

\end{document}